\documentstyle[preprint,aps]{revtex}
\begin{document}
\input epsf
\epsfverbosetrue
\title {Dense, Strongly-Interacting Matter:\\
Strangeness in Heavy-Ion Reactions 1-10~A~GeV}
\author{C.A. Ogilvie }
\address{Massachusetts Institute of Technology, Cambridge, MA 02139 }
\maketitle
\begin{abstract}
This talk reviews the physics of dense, strongly
interacting matter. The newly acquired experimental data
on strangeness production in heavy-ion reactions
from 1-10 AGeV is used to probe the physics of
in-medium properties of matter, multi-body collisions,
and the search for the baryon-rich quark-gluon plasma.
\end{abstract}

\section{Introduction}
\label{sec:intro}

In 1996 the heavy-ion community explored
a whole new physics landscape by measuring Au+Au reactions
at beam energies of 2, 4, 6 and 8 AGeV.
With experimental data to link the AGS full-energy of 10 AGeV and 
the BEVALAC/SIS energies of 1-2 AGeV, we can
uniquely explore how the physics of strongly-interacting
matter changes as a function of density. 
The results from experiments E895 and E866/E917 @ AGS 
and KaoS @ SIS are becoming available, so
it is timely to review the physics of strange particle production  
across the energy range from SIS to AGS.

There are at least three interconnected physics
questions about how matter of increasing density
behaves; 1) what are the appropriate degrees-of-freedom?,
multiply-colliding hadrons and their excited resonances, 
or is the matter dense enough such that 
a baryon-rich quark-gluon plasma\cite{QM96} is formed, 2) if the system 
remains in the hadronic sector, 
do the hadrons propagate with an effective mass different
than their free mass\cite{Brown}, and 3) do 
multi-body, instead of two-body, hadronic collisions play a role
in the reaction dynamics\cite{Ogilvie98a}?

Before reviewing the data and the status of answers
to these physics questions, I would like to expand on each of
the three topics.

In contrast to the high-temperature QGP where quarks and
gluons are predicted to be quasi-free, the baryon(quantum-number)-rich
plasma
is conjectured to have momentum-space correlations between pairs
of quarks\cite{baryonrich}. This pairing is predicted to produce a
color-superconducting quark-gluon phase of matter. The transitions
between this phase, the quasi-free quark-gluon phase, and normal
hadronic matter 
could make a rich and varied phase diagram of strongly-interacting
matter. However there is currently little
knowledge of what signatures could indicate the onset
of the superconducting quark phase, or whether the standard
battery of high-temperature QGP signatures would retain their sensitivity 
in the baryon-rich phase.

Dense matter, which at several times normal nuclear density 
($\rho\sim 0.6-0.7\rm{fm}^{-3}$) may be 
insufficient to form a quark-gluon plasma,
should still be considered a novel and fascinating system in its own right. 
At four times normal density 
the mean-free path ($\lambda=1/\rho\sigma$)
of typical hadrons is a strikingly small 0.5~fm.
This estimate of the mean free path is smaller
than the typical size of a hadron which implies that
the system is so densely packed that each hadron
can collide several times per 1fm/c.
Such multiple collisions will not be independent
of each other and could behave
differently than a succession of two-body
collisions\cite{Ogilvie98a}.

At considerably lower densities there 
are several predictions\cite{Weise96} that kaons  
have in-medium properties.  
K$^-$ are predicted to interact attractively with
baryons, which can be implicitly accounted
for by either an attractive mean-field\cite{Li96,Ehe96}
or by decreasing the K$^-$ effective mass
with density\cite{Weise96}.
K$^+$ are expected to have a repulsive 
interaction with baryons\cite{Li96,Ehe96}.
In transport calculations of heavy-ion reactions, these in-medium effects
are predicted to
strongly increase the yield of K$^-$ due to the increase of phase-space
from the drop in K$^-$ mass.
The spectra of K$^+$ and K$^-$ are also predicted to be shifted
in opposite directions\cite{Fang93,Li96}, with
K$^+$ repelled to higher transverse momentum and K$^-$ 
attracted to lower values of
transverse momentum.

By changing the density it may be possible, but very difficult, to
identify any density-dependent in-medium effect and separate this
from other processes that affect kaon production,
e.g. multiple secondary collisions.
To establish the existence of in-medium
physics will require the systematic use of the
measured K$^+$ and 
K$^-$ yields and 
spectra as a function of beam-energy. As a
strong consistency check,
any posited explanation of an experimental result
that supports in-medium physics, must also explain
data at other beam energies, as well as
the other observables that are sensitive to in-medium effects. 

By studying strangeness production as a function of
beam energy we can probe these three interlinked physics questions:
in-medium properties, multi-body collisions, and the baryon-rich
QGP.  
This will help us make progress towards a consistent
understanding of the behavior of strongly-interacting matter as a function
of density. 
  
\section{Strange Yields in Reactions 1-10 A GeV}
\label{sec:yields}

In Figure 1 the yield of K$^+$ per projectile
participant is plotted versus the number of
projectile participants in Au+Au reactions
at the full AGS energy of 11.6AGeV\cite{ahle98}. The number of participants
is calculated from the measured energy of the projectile spectators.  
The yield per projectile participant in central reactions
is 3.5 times larger than the 
K$^+$ yields from the isospin-averaged N-N collisions
at the same beam energy. 
This extra production steadily increases with the number of projectile 
participants and hence most probably comes from multiple 
secondary collisions of
hadrons in a heavy-ion reaction\cite{Sorge90,Pang92,Li95}.
This mechanism is aided by
many of the secondary collisions being between one or two hadronic
resonances with the energy of the resonances available for kaon
production.

In Au+Au collisions at 1 AGeV\cite{auau1gev}, K$^+$ production
per participant also increases with centrality (Figure 2), 
suggesting that a similar secondary collision
production mechanism may work at both SIS and AGS energies.
The pion yield per participant (Figure 2)
is nearly independent of centrality. It is possible that
the difference between pions and kaons is the strong
absorption of pions.

The kaon yield can be parameterized 
as a power law in the number of participants
\begin{equation}
\rm{yield}=a\times N_{pp}^\alpha
\end{equation}
At SPS energies, where the centrality dependence
of kaons yields is not yet available, the power-law
exponent for $\Lambda$ production is used instead\cite{WA97}.
The power-law exponent $\alpha$ decreases with higher beam 
energies (Figure 3).
If kaon enhancement is defined as the yield in central collisions
compared to either peripheral collisions, or N-N reactions,
then the enhancement of strangeness is largest at the lower
beam energies. 
This is consistent with
secondary collisions  
increasing in relative 
importance for kaon production
compared to initial collisions as the beam energy is reduced.  

The change in the power-law exponent for K$^+$
production demonstrates that
heavy-ion yields have a different beam energy dependence
than proton-proton reactions. Secondary collisions
will also change the energy dependence
of the K$^-$ yield. This is an
effect which will need to be firmly
understood before any in-medium mass reduction of K$^-$
can be inferred from the yield of K$^-$ or the shape
of its excitation function\cite{gsikaon}.

It is instructive to
compare K$^-$ and K$^+$ production by plotting the ratio
of total yields as a function of the number of participants.
Figure 4 shows the K$^-$/K$^+$ ratio in Au+Au
reactions at 11.6 AGeV/c\cite{ahle98}, and figure 5 plots
this ratio for Ni+Ni reactions at 1.8 AGeV\cite{gsikaon}. 
At both beam energies, and also
at the intermediate energy of 4 AGeV,
the K$^-$/K$^+$ ratio is independent of centrality. 
Whatever mechanism drives
the enhancement of kaons with centrality, 
it similarly affects the production of both K$^+$ and K$^-$.

Given the many differences between  K$^-$ and K$^+$ production,
absorption, and possible opposite
changes of in-medium mass, 
the observation that the  K$^-$/K$^+$ ratio 
is constant with centrality is a puzzle. 
At each beam energy, the volume of dense matter, and even
the density should increase with centrality. 
Any possible K$^-$ mass decrease with density should lead
to an increase in the K$^-$/K$^+$ ratio with centrality.
This is not observed, potentially placing a strong constraint
on the existence of in-medium kaon properties.

A review of the theoretical effort to model these reactions
is outside the scope of this paper. For our purposes we can divide
the models into two classes, those that include the possibility
of hadrons having an effective in-medium mass (HSD\cite{Ehe96},
RBUU\cite{Li}, ART\cite{Li95}) and those that use normal masses
throughout (RQMD\cite{Sorge90}, UrQMD\cite{Bass} and the above codes
run in normal-mass mode).

Calculations that use in-medium masses successfully  
reproduce the experimental yield of K$^-$ to within ~30\% 
at both GSI\cite{Li,Geiss} and AGS\cite{Geiss} energies.
The K$^-$/K$^+$ ratio has also been studied\cite{Li}.
The calculated ratio increases with centrality
when in-medium masses are used, however the predicted effect is
negated by the increased absorption of
K$^-$ in dense matter.
While such a cancelation is plausible, it would be
unusual for the cancelation to occur at all energies from 1 to 10AGeV
and as such, the explanation should be checked across this
full energy range.

No consistent picture has yet to emerge
on the calculated K$^+$ yields. Using  an increasing 
in-medium mass for K$^+$, transport 
calculations underpredict the measured yield by factors of two\cite{Geiss}. 
As a cross-check on the consistency of
K$^+$ production in these models, 
RQMD, HSD, and ART calculations with unmodified,
normal masses
for Au+Au reactions at 11.6 AGeV/c are compared in
figure 6.
RQMD impressively reproduces the centrality
dependence of kaon production, whereas both
ART and HSD, for which only central calculations
are available, produce too few kaons. 
These calculations nominally include the
same physics, i.e. multiple scattering and
excitation of a range of resonances, yet for central
collisions there is a factor of two discrepancy between the 
predictions.
This indicates the current level of model
uncertainty of how to calculate the details of a cascade
of interacting hadrons. 
Given this range of model systematics it is premature
to claim the existence of in-medium masses using a model
comparison of kaon yields\cite{Li,Geiss}.

The situation can be improved by constraining the 
models to reproduce the high precision p+p$\rightarrow$K$^+$ data
from COSY\cite{cosy}, and the semi-inclusive
p+A data at AGS energies that is becoming available.
As an example the yield of K$^0_s$ in p+Au at 18 GeV/c
from E910\cite{cole} is shown in Figure 7. The yield increases
rapidly with the number of slow particles associated
with the reaction, consistent with the hypothesis
that as the proton suffers multiple collisions, it
increases both strangeness and the overall yield
of particles. Successful transport models should be able to
reproduce the wide range of p+p, p+A reactions
as an entry step towards the harder problems of A+A reactions. 

\section{Kaon Spectra}
\label{sec:spectra}

Insight into the dynamics of particle production
and propagation
can be obtained by examining the spectra of kaons. 
In the top panels of Figure~8, dN/dy distributions 
for K$^+$ and K$^-$ are shown for a mid-central event class
of Au+Au reactions at 11.6 AGeV/c\cite{ahle98}. 
These have been extracted from exponential
fits to invariant spectra as a function
of transverse mass.
Also extracted from each transverse spectrum is
the inverse slope,  T. The rapidity dependence
of the inverse slope parameter
is shown in the lower panels of
Figure~8.
The dN/dy distribution for K$^-$ 
is narrower than for K$^+$, 
and the K$^+$ inverse 
slopes tend to be slightly larger than 
the inverse slope for K$^-$.  
Both observations are consistent with
there being
less phase space available for K$^-$ 
due to the higher energy threshold for K$^-$K$^+$  pair production
compared to K$^+\Lambda$ associated production.

In contrast, the transport model RQMD 
predicts that the K$^-$ inverse slopes are larger than those for K$^+$.
This can be seen in Figure 9 which plots
the difference of slopes,\\ T$_{K+}$-T$_{K-}$.
The incorrect ordering of slopes in RQMD could be caused 
by a too strong 
absorption of low m$_t$ K$^-$ effectively
increasing the K$^-$ inverse slope.
This could merely be a problem with the
parameterization or 
alternatively it could indicate the need
to include either multi-body
collisions\cite{Ogilvie98a}
or an attractive in-medium potential to the transport model.
This would effectively soften the K$^-$ spectra
as has been observed in the in-medium HSD calculations 
for Ni+Ni reactions at 1.8 AGeV\cite{Geiss}. 

In addition to the softening of spectra, a rather
striking low-p$_t$ rise has been predicted
in the K$^-$ spectra due to the attractive mean-field\cite{Li}. 
In preliminary E864 data\cite{manhat} at the AGS there is no observation
of such a rise even at values of m$_t$-m$_0$=50 MeV. At GSI
the predicted rise occurs just beyond the current experimental 
acceptance\cite{Li}.

One sensitive probe of the possible existence of in-medium kaon
properties is collective flow. Figure 10
shows the measure K$^+$ squeezeout\cite{Shin},
or preferential emission at azimuthal angles 90$^o$ and 270$^0$
from the event reaction plane, in Au+Au reactions at 1AGeV.
The RBUU calculations that include an in-medium repulsive
kaon potential better reproduce the measured azimuthal anisotropy.

\section{Strangeness as QGP Signature}
\label{sec:qgp}

The 
kaon production rate from a QGP
has been predicted to be larger than the hadronic production 
rate, leading
to the suggestion that strangeness
enhancement is a plausible QGP signature\cite{Raf82}. 
As the beam energy is increased from 1 to 10AGeV,
the density of the participating matter should 
increase\cite{caveats}.
A small but increasing volume of QGP might be observable
as an additional increase in kaon production
beyond the hadronic enhancement from multiple collisions.
This would not necessarily be a sharp onset with
beam energy, but rather
a change of evolution in strange yields.

The mid-rapidity yields of $\pi^+$ and 
K$^+$ as a function of the initial available energy $\sqrt{s}$ are 
shown in the upper panels
of Figure~11.  
Both  
pion and kaon yields increase steadily and smoothly with beam energy.   
There is no statistically significant 
indication of any sudden increase,
or change in evolution of the particle yield with increasing
$\sqrt{s}$.  

In the lower panels of Figure~11, 
the $<m_t>$ of the transverse 
spectra minus the rest mass for
$\pi^+$ and K$^+$ are plotted versus $\sqrt{s}$.
The $<m_t>$ is calculated from each fit to the transverse spectrum
and hence includes an extrapolation to low m$_t$.
Compared to the increase 
of particle production, the $<m_t>$ increases
slowly with beam 
energy. The extra available energy mainly goes into more particle
production rather than increased transverse energy.

The increase in kaon mid-rapidity yield  
with beam energy in Figure 11 is more rapid than the increase of pion yield.  
This is emphasized in 
Figure~12, 
where the $K^+/\pi^+$ ratio of dN/dy is plotted versus $\sqrt{s}$.  
This ratio increases steadily from 
near 3\% at 2 A~GeV to 19\% at 
10.7 A~GeV. 
The mid-rapidity K$^+$/$\pi^+$ ratio
from central Au+Au reactions at 1~A~GeV from the KaoS
collaboration\cite{auau1gev} is K$^+$/$\pi^+=(3\pm 1)\times 10^{-3}$.
This is one order of magnitude below
the K$^+$/$\pi^+$ ratio at 2~A~GeV.                
Also shown in this figure is the  
$K^+/\pi^+$ ratio measured in central Pb+Pb reactions at
CERN\cite{NA49}. 
This ratio is comparable, if somewhat lower, than the AGS result.

The double ratio,
the measured Au+Au K$^+/\pi^+$ ratio
divided by the p+p K$^+/\pi^+$ ratio
is shown in Figure~13.
This double ratio
is greater than one, demonstrating that
K$^+$/$\pi^+$ is enhanced in Au+Au 
reactions relative to p+p collisions.  
This enhancement steadily decreases and is low at SPS beam 
energies, possibly because secondary collisions  
decrease in relative 
importance for kaon production
compared to initial collisions as the beam energy is increased.
A steady decrease in enhancement coupled with
the smooth increase in the K/$\pi$ ratio from p+p reactions
naturally produces a maximum in the K/$\pi$ ratio
in A+A reactions.
        
\section{Upcoming Results}

Many new measurements on strangeness have recently been made.
At GSI the KaoS collaboration is analysing proton and carbon-induced 
reactions to provide a baseline for kaon production via
multiple secondary collisions.
Detailed K$^+$ and K$^-$
spectra at 4 AGeV from E866@AGS will
provide more information on possible in-medium effects.
Complementing the kaon data is the
measurement of the
$\Lambda$ excitation function by E895@AGS.
Multi-strange production from Au+Au (E896) and
p+A reactions (E910) at the AGS will be soon
available to compare to the results at SPS.
In addition E917@AGS is analysing
the centrality dependence of anti-lambda
as well as
the mass, width of the decay $\phi\rightarrow$K$^+$K$^-$.
 
\section{Conclusions}
\label{sec:conc}

The availability of heavy-ion data at beam energies of 2, 4, 6, 8 AGeV
has bridged the gap between BEVALAC/SIS and  AGS accelerators.
This provides the opportunity to systematically study the
properties of strongly-interacting matter as a function
of the density. Two intriguing possibilities are the
whether hadrons change their properties in a dense medium,
or whether the baryon-rich quark-gluon plasma
can be found.

The observation of K$^+$ squeezeout 
in Au+Au reactions at 1 AGeV is consistent with the
presence of a repulsive kaon potential. However 
transport models
that include such effects underpredict the yield
of K$^+$. In-medium effects are also predicted
to increase the yield of K$^-$, potentially changing
the K$^-$/K$^+$ ratio as a function of centrality
and causing a sharp low p$_t$ rise in the K$^-$ spectra. Neither effect
is observed.

Taking the above data together as a whole, there is no
consistent evidence for kaon in-medium properties.
In addition using model comparisons to extract this physics
is difficult.   
The factor of two discrepancy between different models with nominally
the same physics input indicates the current level of 
uncertainty in how to calculate a cascade
of interacting hadrons. 
Given this range of model systematics it is premature
to claim the existence of in-medium masses using a model
comparison of kaon yields.

At densities four to five times
normal nuclear matter densities multi-body (as opposed
to two-body) collisions could dominate the dynamics.
This topic needs more attention, in particular how to
distinguish multi-body collisions from a succession of
two-body interactions.

The measured particle yields and the K$^+$/$\pi^+$ ratio
smoothly evolve with beam energy between 1 and 10 AGeV.
There is no indication of a change in behavior that could indicate
the onset of a baryon-rich QGP.
By establishing an energy dependence of particle
production, the data provide a stronger link between AGS
and SPS energies.
Interpolating between the newly available 
excitation function and the full SPS energy leads to
empirical predictions for quantities such as
the K/$\pi$ ratio at energies of 30-40 AGeV, the planned
low-energy run at the SPS.
The extension of the excitation function from 
SIS$\rightarrow$AGS$\rightarrow$SPS will provide a new opportunity
to examine  how strongly interacting matter evolves with density
and whether there is any indication of the onset of
the quark-gluon plasma.

\begin{figure}[htbp]
\begin{center}
\epsfxsize=12.0cm\epsfbox{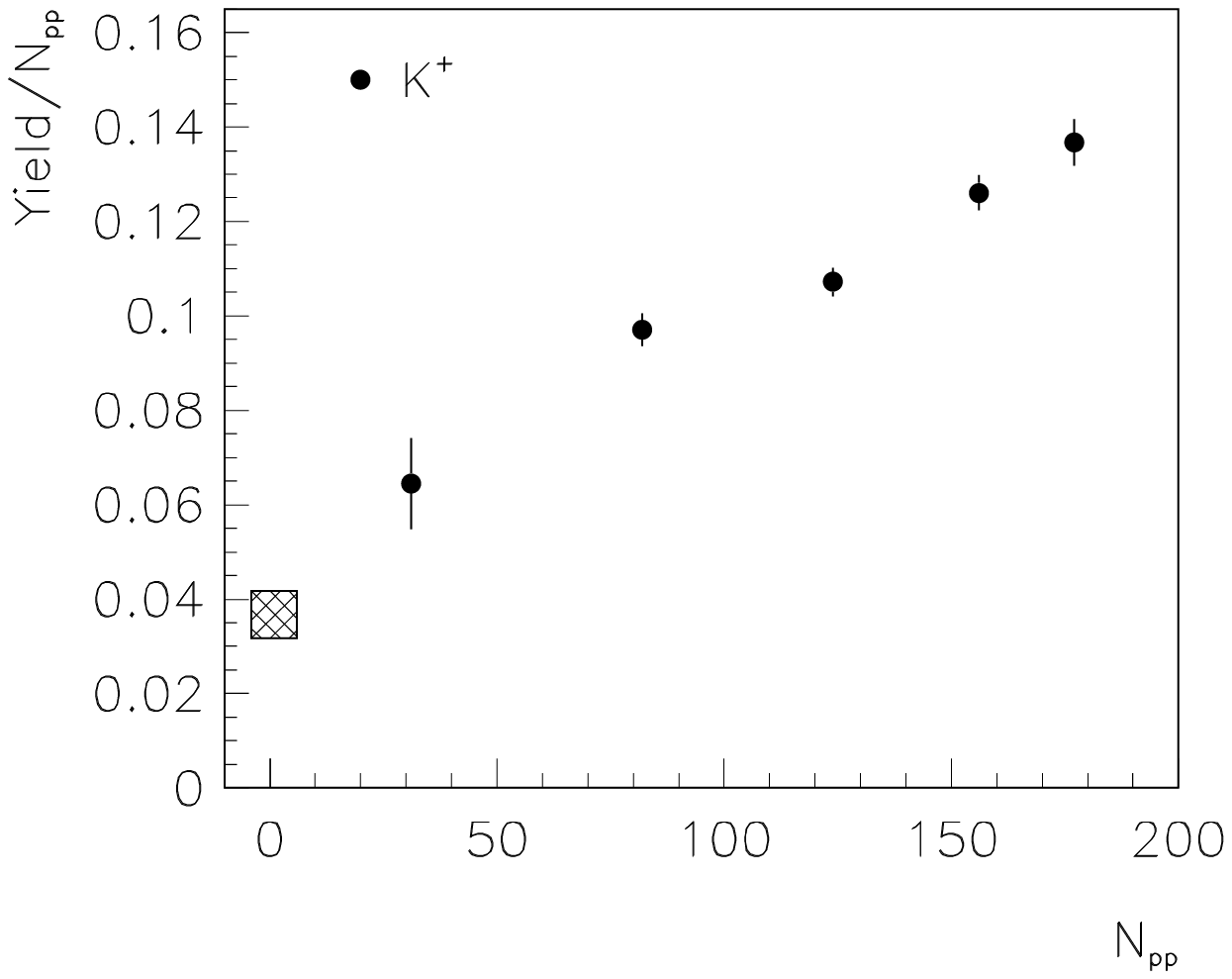}
\caption{The total yields of K$^+$ per projectile participant 
versus the
number of projectile participants, N$_{pp}$,
in Au+Au reactions at 11.6 A~GeV/c measured by E866\protect\cite{ahle98}.
The box on the left is an estimate of the kaon yield from
the initial N+N collisions.}
\end{center}
\label{fig:agskper}
\end{figure}

\begin{figure}[htbp]
\begin{center}
\epsfxsize=12.cm\epsfbox{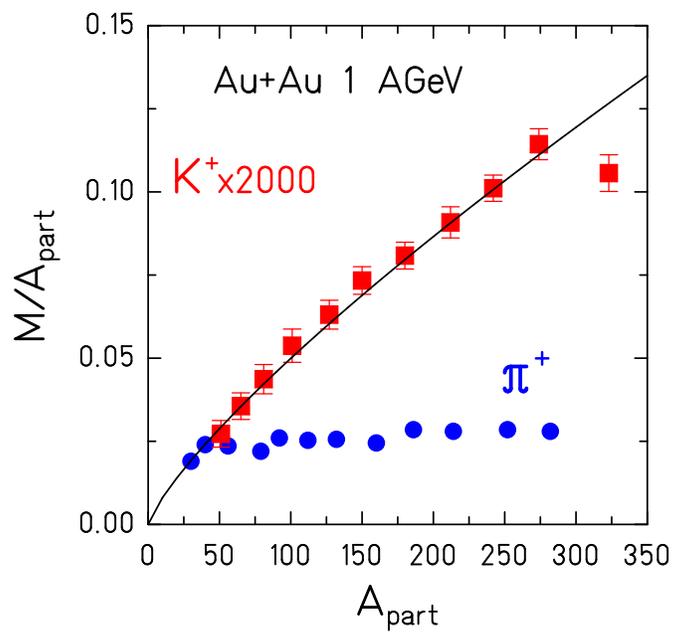} 
\caption{The total yields of K$^+$ and $\pi^+$ per participant 
versus the
number of participants, A$_{part}$,
in Au+Au reactions at 1 A~GeV measured by KaoS\protect\cite{auau1gev}.}
\end{center}
\label{fig:gsikper}
\end{figure}

\begin{figure}[htbp]
\begin{center}
\parbox{8.0cm}{
\epsfxsize=12.0cm\epsfbox{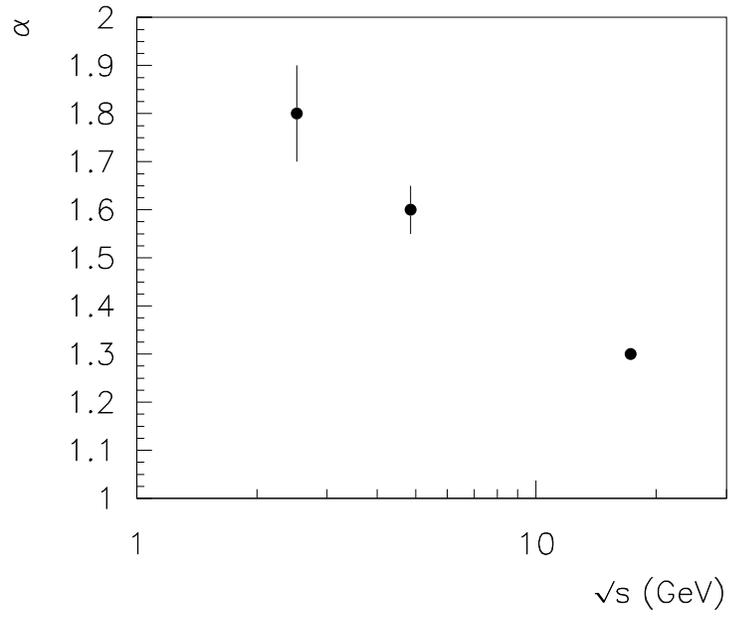}}
\parbox{8.0cm}{ 
\caption{The beam energy dependence of the power-law
exponent $\alpha$ (equation 1) that describes the
non-linear increase of strangeness production with
centrality.
}}
\end{center}
\label{fig:alpha}
\end{figure}

\begin{figure}[htbp]
\begin{center}
\parbox{6.0cm}{
\epsfxsize=12.0cm\epsfbox{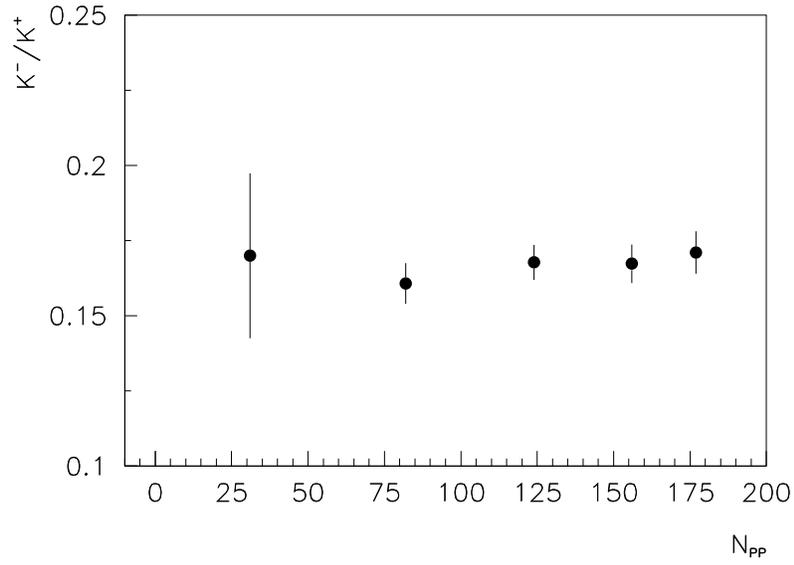}}
\caption{The K$^-$/K$^+$ ratio of total yields 
versus the
number of projectile participants, N$_{pp}$,
in Au+Au reactions at 11.6 A~GeV/c measured by E866\protect\cite{ahle98}.} 
\end{center}
\label{fig:agsrat}
\end{figure}

\begin{figure}[htpb]
\begin{center}
\parbox{6.0cm}{
\epsfxsize=12.0cm\epsfbox{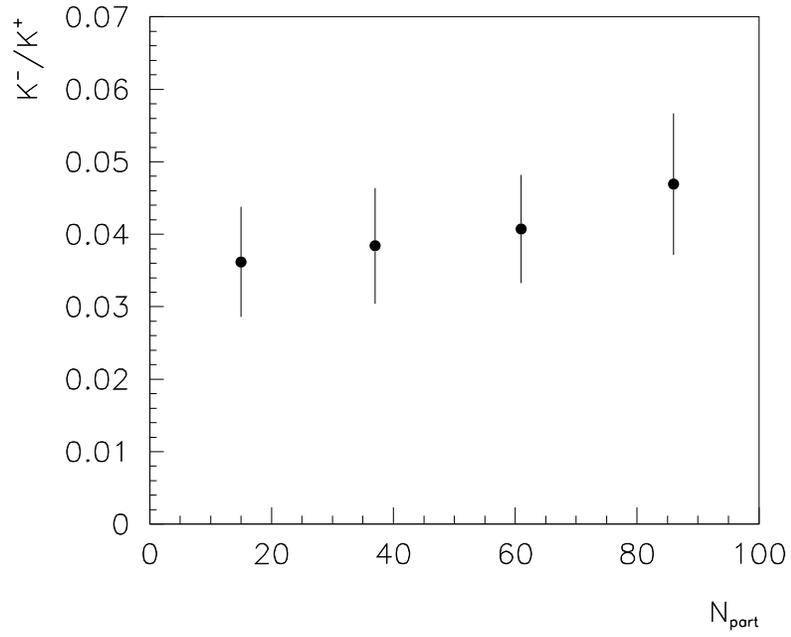}}
\end{center}
\parbox{12.0cm}{ 
\caption{The K$^-$/K$^+$ ratio of total yields 
versus the
number of participants
in Ni+Ni reactions at 1.8 A~GeV measured by KaoS\protect\cite{gsikaon}.
}} 
\label{fig:gsirat}
\end{figure}

\begin{figure}[htbp]
\begin{center}
\parbox{8.0cm}{
\epsfxsize=12.0cm\epsfbox{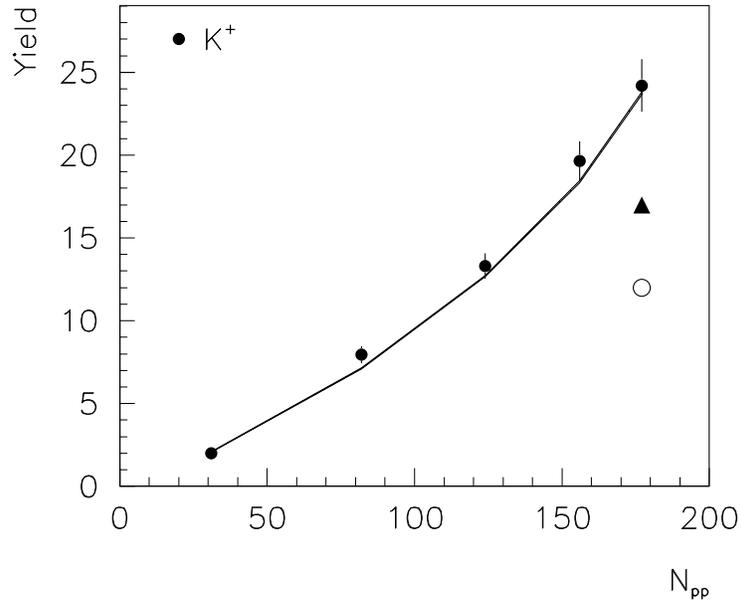}}
\parbox{8.0cm}{ 
\caption{The measured K$^+$ 
yield in Au+Au reactions at 11.6~A~GeV/c
versus the number of projectile participants\protect\cite{ahle98}.
The data are compared to the hadronic transport
model RQMD (v2.3)\protect\cite{Sorge90} (solid line)
and the HSD\protect\cite{Ehe96} (open circle) and ART\protect\cite{Li96}
(filled triangle)
models run with normal hadronic masses.}}
\end{center}
\label{fig:kpmodel}
\end{figure}

\begin{figure}[htbp]
\begin{center}
\parbox{8.0cm}{
\epsfxsize=12.0cm\epsfbox{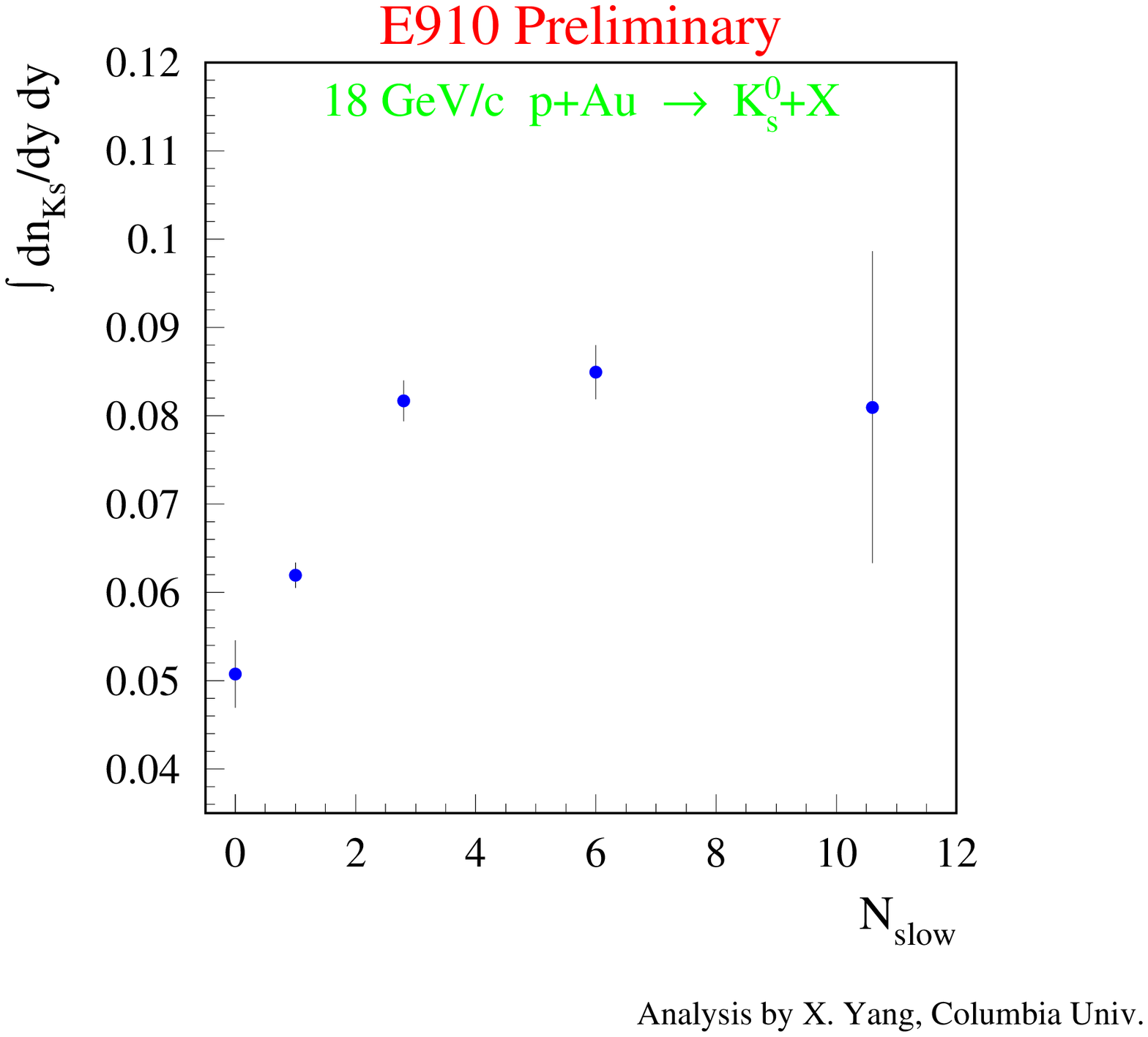}}
\parbox{8.0cm}{ 
\caption{The preliminary integrated K$_s^0$ yield in p+Au reactions at
18 GeV/c versus the number of slow particles
emitted in the event as measured by E910\protect\cite{cole}.
}}
\end{center}
\label{fig:e910}
\end{figure}

\begin{figure}[htpb]
\begin{center}
\parbox{8.0cm}{
\epsfxsize=12.0cm\epsfbox{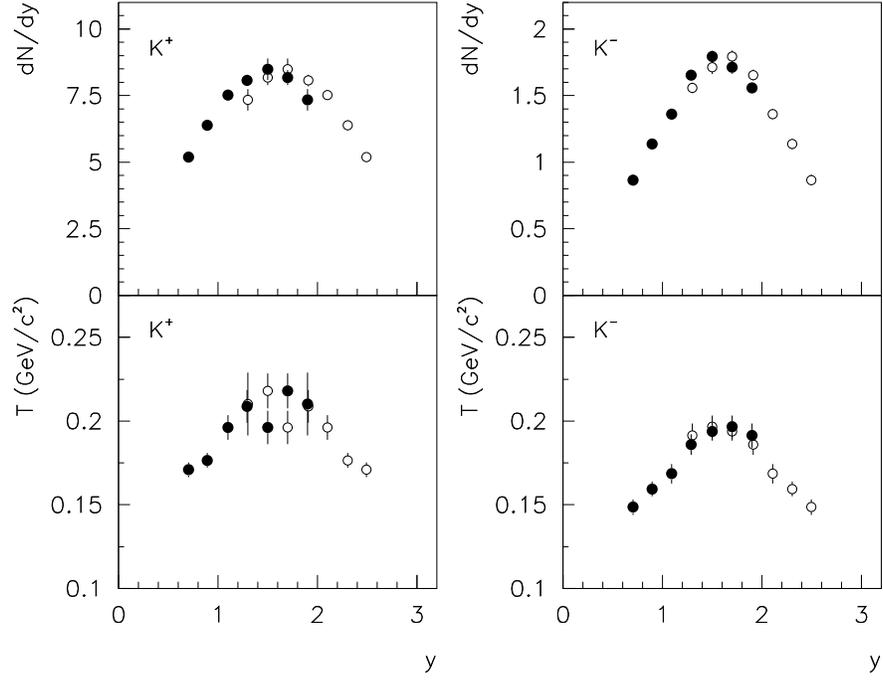}}
\parbox{8.0cm}{ 
\caption{The dN/dy and inverse slope
distributions as a function of rapidity for
K$^+$ and K$^-$ emitted from mid-central
Au+Au reactions at 11.6 A~GeV/c\protect\cite{ahle98}.
The hollow circles are the data points reflected
about mid-rapidity (y=1.6).}
}
\end{center}
\label{fig:kdndy}
\end{figure}

\begin{figure}[htpb]
\begin{center}
\parbox{8.0cm}{
\epsfxsize=12.0cm\epsfbox{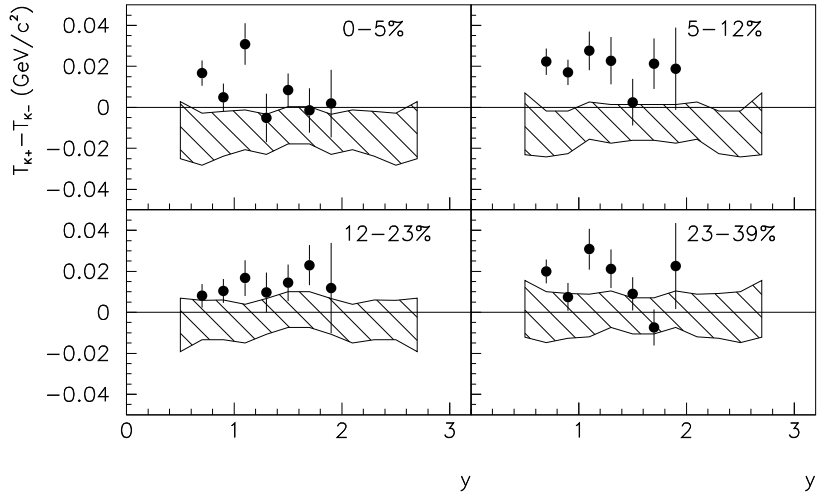}}
\parbox{8.0cm}{ 
\caption{The difference in inverse slope parameter 
T$_{K^+}-\rm{T}_{K^-}$ versus y for 
the different
centrality classes 
in Au+Au reactions at 11.6 A~GeV/c.
The centrality classes are categorized by the \%
of total reaction cross-section (6.8 b)\protect\cite{ahle98}.
The data are compared to the hadronic transport
model RQMD (v2.3)\protect\cite{Sorge90} .
The dashed band is centered at the
model prediction and has total width of $\pm 1\sigma$.
}}
\end{center}
\label{fig:tdiff}
\end{figure}

\begin{figure}[htbp]
\begin{center}
\parbox{8.0cm}{
\epsfxsize=12.0cm\epsfbox{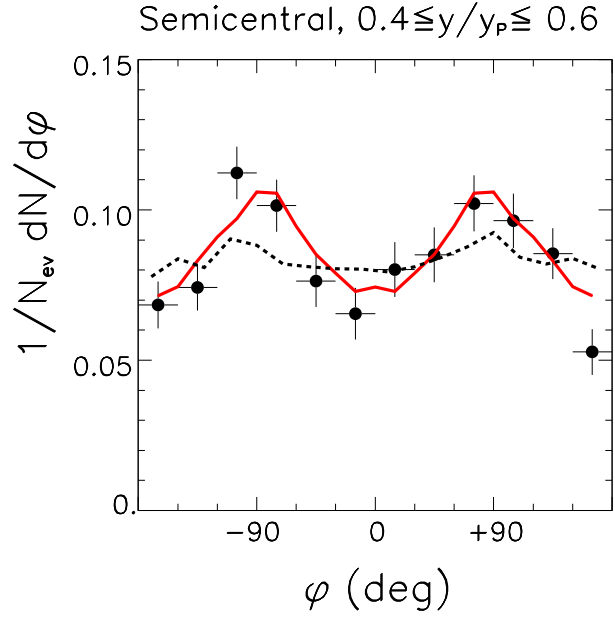}}
\parbox{8.0cm}{ 
\caption{The azimuthal distribution of K$^+$ in 
semi-central Au+Au
collisions at 1 AGeV as measured by KaoS\protect\cite{Shin}.
The lines represent RBUU calculations for an impact
parameter of b=7fm without (dashed line) and
with an in-medium KN potential (solid line).
}}
\end{center}
\label{fig:squeeze}
\end{figure}

\begin{figure}[htbp]
\begin{center}
\parbox{8.0cm}{
\epsfxsize=12.0cm\epsfbox{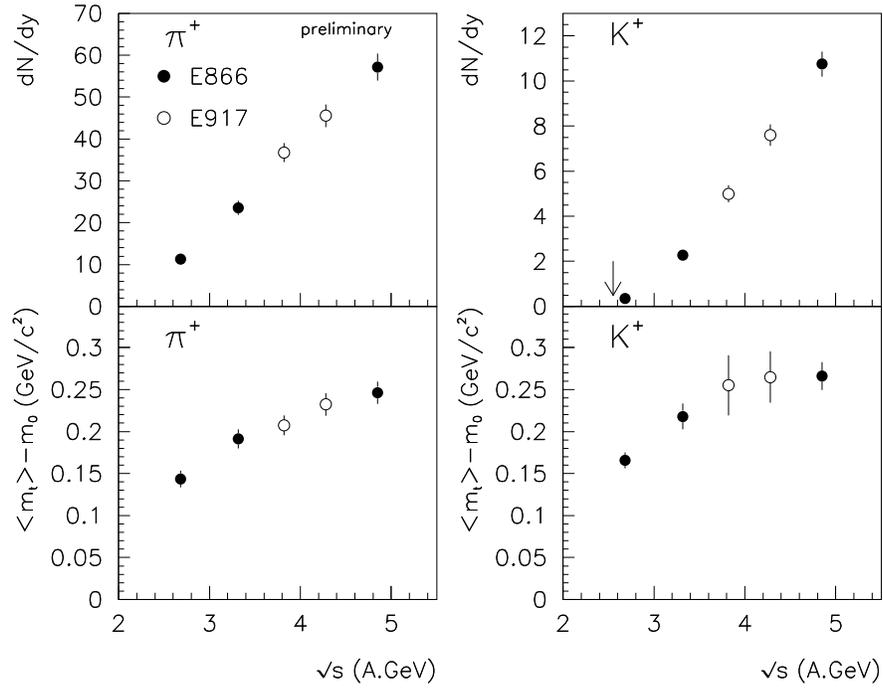}}
\parbox{8.0cm}{ 
\caption{The yield of $\pi^+$ and K$^+$ at mid-rapidity (top-panels)
for central Au+Au
reactions as a function of the initial available beam energy.
The lower panels
show the mean m$_t$ minus the rest mass
for $\pi^+$ and K$^+$ at 
the same rapidity.  
}}
\end{center}
\label{fig:excit}
\end{figure}

\begin{figure}[htbp]
\begin{center}
\parbox{6.0cm}{
\epsfxsize=12.0cm\epsfbox{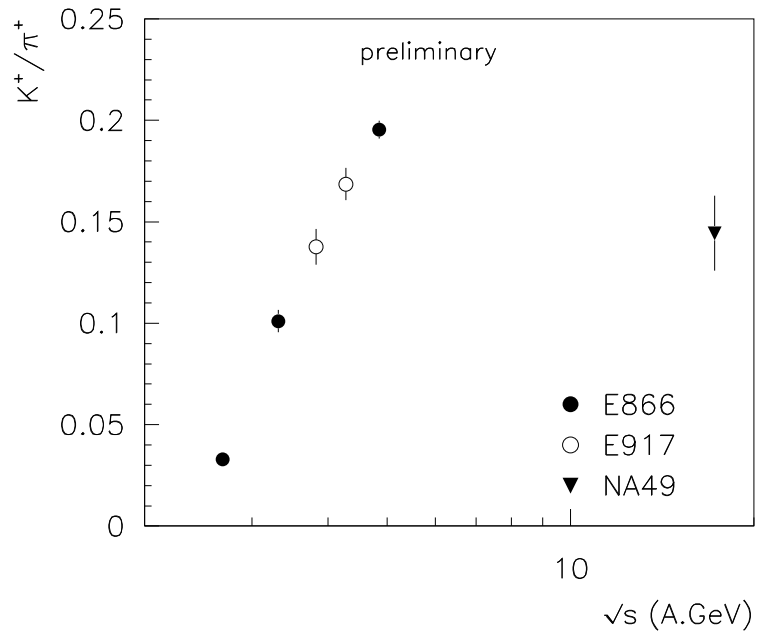}}
\end{center}
\caption{The ratio K$^+$/$\pi^+$ at mid-rapidity in
central Au+Au reactions as a function 
of the initial available energy.}
\label{fig:kpi}
\end{figure}

\begin{figure}[t]
\begin{center}
\parbox{6.0cm}{
\epsfxsize=12.0cm\epsfbox{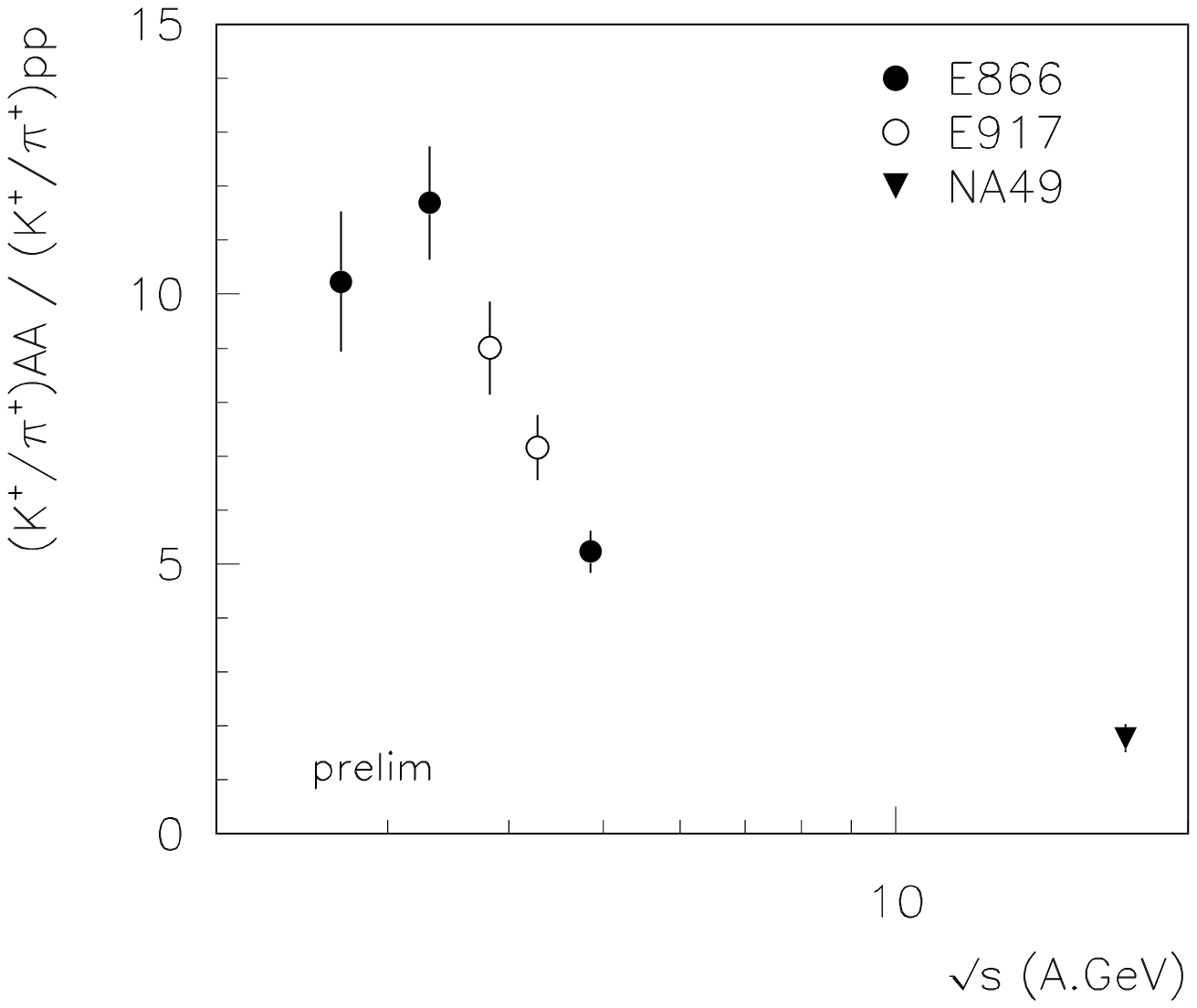}}
\end{center}
\parbox{10.0cm}{ 
\caption{The double ratio K$^+$/$\pi^+$ at mid-rapidity from central
Au+Au reactions divided by K$^+$/$\pi^+$ of total yields from p+p
reactions as a function 
of the initial available energy. 
}}
\label{fig:enhance}
\end{figure}


\begin{thebibliography}{99}
\bibitem{caveats} There is currently no proof that the baryon-density
of participant matter increases monotonically between 1 and 10 AGeV.
At some beam energy it is expected that a maximum density
is reached.
\end{thebibliography}
\end{document}